# Scientific Collaboratories as Socio-Technical Interaction Networks: A Theoretical Approach

**Rob Kling, Geoffrey McKim, Joanna Fortuna, and Adam King**
Indiana University School of Library and Information Science
10[th] & Jordan, Bloomington, IN 47405 USA
+1 812 855 5113
kling@indiana.edu

**ABSTRACT**
Collaboratories refer to laboratories where scientists can work together while they are in distant locations from each other and from key equipment. They have captured the interest both of CSCW researchers and of science funders who wish to optimize the use of rare scientific equipment and expertise. We examine the kind of CSCW conceptions that help us best understand the character of working relationships in these scientific collaboratories. Our model, inspired by actor-network theory, considers technologies as Socio-technical Interaction Networks (STINs). This model provides a rich understanding of the scientific collaboratories, and also a more complete understanding of the conditions and activities that support collaborative work in them. We illustrate the significance of STIN models with several cases drawn from the fields of high energy physics and materials science.

**Keywords**
Socio-technical networks, actor-networks, collaboratories

## INTRODUCTION
We are in the midst of a *revolution about the expectations* of how IT can substantially improve communications and collaboration among scientists, as well as with professionals and broader publics. From the beginnings of the Internet, funding for IT infrastructure is frequently justified in terms of speeding up and widening access to scientific communication. Many of the expectations are based on conceptions of high speed telecommunications enabling information to move rapidly and relatively inexpensively "anywhere anytime" – thus enabling low cost and widely available electronic journals, preprint servers, collaboratories and so on.

These expectations have both fostered and conditioned the development of a variety of new scientific collaboratories in fields such as upper atmospheric physics [28], environmental biology [23], and solid surface physics [34]. The term "collaboratory" is often used to refer to laboratories where scientists can work together while they are in distant locations from each other and from key equipment [3]. However, some collaboratories are not used to the degree and in the manner in which they were intended.

The character of the working relationships in these collaboratories are strongly shaped not only by social relationships – such as those between project managers and employees, scientists and technicians, but also by relationships between actors and technologies. For example, extensive and deep expertise may make a scientist who designed a collaboratory into a desirable collaborator for cutting edge research that uses her equipment. Scientists may be constrained in their ability to make effective use of a collaboratory by the tool sets in use at their institutions.

In this article, we present a theoretical model that will help understand the **character of working relationships, both during development and during routine operations of a scientific collaboratory**. We find that this theoretical model, Socio-Technical Interaction Networks (STINs), provides a richer understanding of the scientific communications collaboratories, and also a more complete understanding of the conditions and activities that enhance the sustainability of a communications forum within a field. We will then illustrate the usefulness of STINs through two case studies – one in high-energy physics and one in materials science. Socio-technical network models are specially interesting for CSCW analysts and designers because they help to theorize technologies as well as social relations. They differ substantially from social theories that have been used to interpret CSCW developments that don't conceptualize technologies, such as situated action, structuration theory and activity theory.

The STIN approach should have high salience to the CSCW community, as it can help explain the sustainability, or conversely, the failure of collaboration within collaborative systems.

### Socio-Technical Interaction Networks
The Socio-Technical Interaction Network approach is inspired by actor-network theory (ANT), as developed in [18], as well as our own prior research about 'web models" of computerization ([16,10,12]). ANT is an ontology that maps out the practice of science and technology in terms of enrollment and mobilization of supporters, and translations of interest in favor of a particular scientific claim or technology. It also focuses on the encapsulation of scientific claims in technologies and instrumentation, and their subsequent use in developing other scientific claims.

The advantage of ANT for CSCW is that it provides insights about the feasibility and sustainability of particular scientific collaboratories. ANT has seen some limited uptake in the CSCW community in analyzing the sustainability of PD projects (see [2,7]). It has two primary weaknesses, from a CSCW perspective: (a) it is more useful in analyzing development of new systems than it is in analyzing routine operations or use, in which explicit mobilization and enrollment are less important; and (b) it provides little guidance on how to draw the networks; it doesn't identify which kinds of enrollments matter most.

The STIN approach is an attempt to correct for these weaknesses. The following brief thought-experiment will illustrate the potential usefulness of the STIN approach in a CSCW context. After that, we will develop the STIN concepts more explicitly.

Imagine that an Indiana University Program in CSCW, as a campus-wide program, has graduate students whose primary affiliations are in departments all over the campus. The Program's faculty want to develop a more active intellectual community in CSCW by encouraging communication and community between these graduate students. In order to foster a stronger sense of community between these geographically dispersed students, they consider several options: (a) create an e-mail listserv for the students to use; (b) create a Web discussion board for the students; (c) create a bi-weekly face-to-face meeting/discussion group with lunch provided; or (d) create a dissertation support group for PhD students currently dissertating. ANT can't tell us anything about the relative value of these alternatives in energizing the students.

The STIN approach can help understand how each of these scenarios might play out, because we can characterize some of the likely interactions based on our prior-knowledge of graduate student groups and the communicative properties of different e-media. For example, the e-mail list may be easy to access remotely. Further most students are in the habit of accessing their e-mail frequently, allowing e-mail to support more interactive communication. However, e-mail has the disadvantage that messages from the discussion list will arrive interspersed with other messages for the student, and thus may have a lower claim on the student's attention than more immediately pressing messages. The Web board has the advantage that communication can easily be separated from other electronic communications the student might receive. However, the students must remember to log into the Web-board. Further, if a student logs into the Web-board and does not see any postings, they may take longer to log in next time. This vicious circle may culminate in non-use.

Both of these scenarios use technologies that inherently generate textual traces of the conversations. This may seem to be an advantage of the technology, until one considers that a frequent topic of conversation among PhD students is the faculty they work with. Many students may be reticent to communicate candidly with such a potential for surveillance, leading to a more formal and less community-building forum.

**Collaboratories as Collections of Instruments vs. Collaboratories as Socio-Technical Networks**

We can illustrate these abstract concepts with different approaches to collaboratories. Wulf [33] defined "collaboratory" as "a center without walls, in which users can perform their research without regard to geographical location—interacting with colleagues, accessing instrumentation, sharing data and computational resources, and accessing information in digital libraries. The "Layer Cake" Model of collaboratories treats them as collections of scientific instruments and information technologies to enable using them and to support collaboration by people who are not co-located. The collaboratory is composed of a set of technologies; the sociality of collaboratories comes from the (collaborative) interaction of "the users" with each other. The layer cake metaphor refers to a gathering of people at a party where the food, such as a layer cake, is a set of material objects; sociality begins when the partygoers arrive and interact with each other over the food.

In contrast, Myers [23] characterizes his Environmental Molecular Sciences Laboratory as one in which scientists who wish to use it have to understand the instrumentation, learn how to prepare samples for it, learn how to use it, and perhaps have the instruments reconfigured for their studies. This understanding and learning requires help from scientists who have significant responsibility for selecting, configuring and maintaining specific instruments. In Wulf's image, there is no one "in the collaboratory" before its users arrive and after they leave. In Myers' account, however, each major instrument has a scientist at its side before "users" come and after they leave. Further, in order to utilize instruments in a collaboratory, a scientist (or team) at a remote location have to develop social relationships, such as trust, with the scientists who know the instruments and who can be viewed as 'inside' the collaboratory.

**METHODS**

We used two methods in performing the research for this article: documentary interpretation and semi-structured participant interviews. First, the research team read exhaustively documentary materials about scientific communications forums in general, and collaboratories in specific.

One of the things we discovered in trying to perform this research has important methodological implications. Despite the wealth of documentary information available, and the availability of these communications forums on the Web, we were unable to learn about many important facets of the scientific communication forums – in particular, the business models and the institutional linkages needed to maintain these systems. These elements turned out to be the crucial pieces in our sociotechnical network-based model of field compendia.

The research team conducted in-depth semi-structured interviews with some of the key shapers of these communications forums. We interviewed shapers of particle physics collaboration web-sites at HEPLAB and a materials science collaboratory, MatterLab, (as well as other scientific forums), from March 1998 through November 1999. In these interviews, we probed issues such as: support and funding for the forum, governance structures, audiences (targeted and actual), the role of the forum in the communication system of the field, assessment of the usefulness or success of the forum, and the opportunities and pressures that lead to new features.

## SOCIO-TECHNICAL NETWORKS

### Socio-Technical Network Models of Collaboratories

Conventional theories of technologies portray them as tools whose adoption by organizations is based on norms of rationality and technical efficiencies. Different ways of configuring technologies in practice are of relatively minor significance. In the case of scientific communication forums, the conventional analyses emphasize the rapidly increasing price/performance of computer hardware, the declining size and weight of equipment, the ubiquity of telecommunications to help people to readily move data readily within and across organizations. The conventional theories tilt towards economic and technological determinisms. For example, some scientists believe that the physics working article (e-print) server at Los Alamos National Labs (Arxiv.org) is the model of publishing that will sooner or later be followed by all of the sciences: it is "just a matter of time [15]". Careful empirical research studies about scientific communication forums and other information technologies have found that "almost identical technologies" are often configured very differently in practice. It is common for preexisting social arrangements to influence these configurations. A "social shaping of technology" perspective suggests caution in trusting deterministic claims. In addition, each social group may have to locally configure scientific communication forums to use them most effectively. What are claimed as "best practices" may work well in some organizations but not others. Thus local R&D costs can remain relatively high and the overall costs of using new scientific communication forums may not fall rapidly. There are important economic and social consequences in the differences between these kinds of predictions.

These theoretical differences are of major practical consequence. In the case of scientific communication forums (broadly), the conventional theories lead us to emphasize the rapidly increasing price/performance of hardware and to anticipate media convergence. Some go farther and "believe that the paper document is dead; we are just not aware of it yet. [33]" Further, one can expect that a few well-crafted pilot projects – done almost anywhere -- can help to establish "best practices" that everyone else can follow. A first stage of social learning about new scientific communication forums can be exploratory and costly; however, subsequent uses elsewhere can be imitative and relatively inexpensive.

### Limited Socio-Technical Conceptions of Collaboratories

Some analysts have been using the term "socio-technical" informally to understand collaboratories and other IT applications. There are two common uses which differ considerable from our own use. The first is that IT applications are technologies that have social consequences. Technologists, such as computer scientists build the IT applications; social scientists then study their consequences for work, organizational forms and other social behavior. A recent "socio-technical summit" about Internet2 was organized on this conception of socio-technical. We will show how the concept of Socio-technical Interaction Networks (STINs) can be put to better use than this.

A second common use is reflected in some of the discussion of collaboratories [26]. In this view, scientific communication forums generally, and collaboratories in particular, can be viewed as layered systems. The bottom layers are various technologies, such as computer networks and specific kinds of applications. The "tool sets" of the collaboratory are the technical layers. The "socio" arises when people use the scientific communication forum to communicate. The behavior of the participants should be understood as "socio-technical" because of the strengths and limitations of the tool sets at any given time. This conception separates "socio" from "technical" by virtue of how the layers are conceptualized. Even so, this conception has undergirded some interesting and important research done under the rubric of Computer Supported Cooperative Work (CSCW) [8]. We refer to this conception as a "layer cake model" in which technologies compose the primary layers and social life abounds between the people who come to party with each other and consume the cake.

### What Are Socio-Technical Interaction Networks?

In our view, the concept of socio-technical behavior should be used to refer to more integrated conceptions of the interaction of people and technologies. In particular, what are referred to as technologies are developed within a social world and supported by technicians and others with specialized skills.

While few scientists have direct experiences with collaboratories, academics are familiar with oral forms of scholarly communication and its alteration by electronic communication. So this makes a good example for explaining one view of Socio-Technical Networks[1]. Amplifiers in lecture halls, video conferencing, and videotape alter the nature of audiences that scholars can reach, and also shift the relationships between those audiences and lecturers/speakers. These electronically enhanced forums do not simply provide "more communication," but also alter the ways that people speak and interact. The speaker may have to work in a special conference room and be separated from local participants by complex equipment (thus altering local interactions). As the audience scales up in size, or moves out in space and time with real-time video or asynchronous-video-tape, the informal give and take between speakers and listeners becomes more difficult (in contrast with the smaller face-to-face seminar). On the other hand, people watching a videotape may privately replay sections to enhance their comprehension, while in a face-to-face meeting they may have to ask questions (that might also embarrass the speaker or questioner).

Voice-based face-to-face conference, video conferencing, and videotape are not simply equipment. They shape scholarly communications as Socio-technical Networks in which social characteristics such as controls over access (via pricing and distribution channels), and social protocols for regulating discussions between speakers and audience also influence character of scholarly communications.

These socio-technical networks are heterogeneous since they bring together different kinds of social and technological elements -- cameramen their cameras, and speakers; editors and their technologies; copyright laws and perhaps even lawyers; funders and their budgets; producers and their time schedules into a 'complex web".

The nature of videotape pricing and the distribution channels can lead to minor or huge expansions beyond the original conferees. Despite scholars' potentially broader access to conference talks via videotape distribution, a face-to-face conference is different from a videotape collection of its talks because of the diverse informal discussions and important social networking that conferences support. The face-to-face conference and the videotape collection are different scholarly communication systems with overlapping capabilities, but which also support very different forms of scholarly communication.

In a similar way, a scholarly journal can also be usefully understood as the product of a socio-technical production and communications network. The publishing communication system includes both full-text materials (articles and books), and indexes/pointers to these materials (including book reviews, abstract sets, specialized bibliographies, and diverse catalogs). The network brings together authors, editors, reviewers, readers, publishing staffs and others.

In addition, the journal is embedded in other socio-technical networks, such as the communication and reward systems of the fields and institutions in which its authors participate and the libraries or archives that store copies, index them, and abstract them. The journal's viability will depend upon how it is positioned within this second network i.e., whether it has the standing to attract high quality authors and readers. The STINs that constitute the journal and in which it participates are directly linked. For example, the editorial board helps to constitute the journal. But the editor's scholarly prestige is one influence on potential authors perception of the journal's quality. This is especially important for new journals [15,16].

The links and nature of the social interactions between various participants and technologies that constitute the Socio-Technical Interaction Network (STIN) can be complex[2]. For example, authors are often in a dependency position relative to the particular editors of a journal who are managing the reviews of their own articles. However, some author-editor relationships can be more complicated if they had prior collegial relationships or prior conflicts. In some cases, the extent to which an author has "hot results" can transform the relationship into one in which the editor tries to court the author, rather than simply administering an independent review of the authors' manuscript. Further, a STIN may include participants who are not necessarily

---

1  We use the term network rather than system because these configurations are open ended and not 'designed.' "A network, by contrast, is loosely organized; often imperfectly integrated; has nodes that may be part of one to many other networks as well; and can be reconfigured. [5]"

2  As in actor network theories (ANT) and web models, social and technical elements are brought together in a network. In the original formulation of ANT, the primary driving social process is one in which some parties try to enlist others in a central project. Law [18] notes that researchers use several different approaches under an ANT rubric. STIN models do not make a committment to a single driving social process. The nature of the relationships and the dynamics of social action are specified explicitly and in addition to specifying the socio-technical network.

Latour [19] provides an intriguing example about the ways that different social theories lead analysts to characterize different participants and their networks. In his example, he feels that a French colleague of his contributed to the drowning of an African informant. However, village elders who investigated the drowning ignored the French, drew very different social networks, and implicated an aunt of the deceased.

direct participants in creating or reading the e-journal, such as the members of Promotion and Tenure committees at the authors' university. They are part of an e-journal's network because authors may decide whether or not to publish in an e-journal based on their expectations of the ways that such committees will (de)value their publications[3].

**Generating the Socio-Technical Interaction Network**

A significant problem faced by sociotechnical analysts is that of how to figure out what belongs in the network and what does not – in other words, how to generate the network. The STIN approach calls out several different social interactions as being generative of sociotechnical networks. These types of social interactions include: resource dependencies and account-taking. Resource dependencies create networks that include groups such as funders and grantees, scientists who develop collaboratories (insiders) and offsite scientists who utilize them (outsiders), employers and employees, and journal publishers, editors, reviewers, and authors. Constructing networks based on resource dependencies highlights several important themes, including the political economy of a forum, various kinds of hidden (articulation) work, and network extension through institutional linkages. Account-taking links an actor to others who serve as "reference points". Scientists may take account of their peers in competing laboratories, the program directors who review their proposals and scientific progress, and the editors and reviewers of conferences and journals who influence the visibility of their research. None of these other scientists may be formal participants in a collaboration; yet they are likely to have some influence on the problems chosen, the ways that they are approached, the instruments used, the pace and scheduling of a collaboratory's work, and the downstream forms of publication (as well as the nature and number of communications between the direct participants in a collaboration).

**"Highly Intertwined" Socio-Technical Interaction Network Models (HISTIN)**

Before we examine some applications of Socio-Technical Interaction Network models of scientific communication forums, it helps to explain one kind of "highly intertwined" Socio-Technical Interaction Network model (HISTIN)[4]. This model seems especially helpful in understanding electronic forums including collaboratories, conferencing systems and electronic journals. The characterization of STINs above separated equipment (or technology) from social relationships and resources.

This analytical separation between artifacts and social worlds is very common, even in social shaping analyses. In these approaches, as in the reinforcement politics theory, social relationships shape the kinds of artifacts selected, their configuration and their typical modes of use. But artifacts are conceptualized as "the products of engineering" and as 100 per cent separable from social relationships.

In the "HISTIN model, technology-in-use and a social world are not seen as separate – they co-constitute each other. The model is "highly" (but not completely) intertwined because its adherents do not insist that this intertwining of technical and social elements is universal. Rather, it is commonplace, and a good heuristic for inquiry, especially with complex technologies. References to technologies and social relations are largely for analytical convenience. For example, one might say that "Indiana University is using web-boards to support class discussions when the participants are not in-class together." Indiana University and its classes would be treated as "social forms" and "web-boards" as material "information technologies." In the "highly intertwined model" the web-boards could be examined to see how they are constituted as socio-technical networks. For example, certain social relationships are inscribed into the web-boards when they are used (such as access controls for who can read or write onto them) and or constituted in their supporting social protocols about legitimate content (to what extent are jokes or advertisements allowed in a specific class's web board?).

Similarly, a social form such as Indiana University in Bloomington can be seen as co-constituted with diverse technologies[5]. Its routine operations rest on a complex set of building technologies, heating/cooling technologies, food acquisition and preparation technologies, and information and communication technologies. Without these (and other technologies) we would have 35,000 students, 1500 faculty and 2000 staff milling around in the forested hills of Bloomington, having tremendous problems in foraging for food, and organizing themselves just by face-to-face conversation, word of mouth, and rumor! In contrast, the Indiana University of 1880 with about 300 students and a few dozen faculty was workable with much simpler technologies than those that are required for the much vaster contemporary university. Any means to record information about enrollments, courses, requirements, etc. would require some kind of scientific communication forums, however crude. In this sense, an organization such as Indiana University is constituted not just of people in social relationships, but also of diverse technologies. In fact, one can interpret many of the discussions of Internet-supported distance education as efforts to constitute new kinds of universities by changing their scientific communication forums infrastructures and pedagogies.

---

[3] The Promotion and Tenure committee in this example is a place-holder for any of the parties who may review an author" publications. An extended diagram might include department chairs and deans who set academic salaries, research grant review committees, etc.

[4] For other accounts that examine socio-technical networks as complexes that intertwine social and technological elements as a complex admixture see [2,20,21,31]

[5] This argument owes much to Strum and Latour, 1999.

The HISTIN model is particularly useful for understanding the social shaping and "consequences" of scientific communication forums which foreground communication between individuals or groups. But even a less restrictive STIN model raises cautions about simple claims about the forum's "impacts" (such as "the Internet is democratizing science").

The STIN Models foregrounds such phenomena as: relations between a collaboration and other scientific teams and actors, content control (setting boundaries around communications and participation), resource dependencies, work required to make a system useful, work and resources required to keep a system sustainable, translations used to mobilize resources, business model and governance structures.

Explicit STIN models have been applied to understanding the IT support scientific research teams (Kling, 1992) and understanding the relative viability of early collaboratories within model organism molecular biology (Star and Ruhleder, 1996). Implicit STIN models have been undergirded studies of IT applications failures (i.e., Markus & Keil, 1994).

Explicit HISTIN models have been applied to understanding the character and development of electronic documents [1] and to the development of Internet standards ([9][24][25]). These studies illustrate that HISTIN concepts are often understood informally in some of the professional IT communities. For example Monteiro [24] studied the processes by which the groups that are responsible for developing Internet infrastructure standards negotiated changes in the underlying address structure in the early 1990s. It is worth noting the language of the official "Request for Comments" for new IP addressing schemes:

"The large and growing installed base of IP systems comprises people, as well as software and machines. The proposal should describe changes in understanding and procedures that are used by the people involved in internetworking. This should include new and/or changes in concepts, terminology, and organization. [29]"

Implicit in this passage is an understanding that the Internet is not simply a technology that is used by people. The "layer cake" model of socio-technical systems, would usually place the Internet as one of the lower layers in a model. In the view of the RFC authors, the Internet is infused with people, and their concepts (of IP addressing), and their procedures (for administering servers), and the various ways that they are organized.

Another illustration comes from Myers' [23] characterization of the Environmental Molecular Sciences Laboratory (EMSL):

"Before deciding which tools to use in their work, researchers first need to consider what occurs when they do science and how collaboration can help. Setting up a collaboratory is not simply a matter of running a remote experiment. Remote control software may let participants perform the experiment, but they will also need access to the sample preparation procedures, instrument settings, and other information usually recorded in a local paper notebook today. Before the experiment can be considered, potential participants must discover the remote resource, understand its capabilities, contact the local researchers, develop trust, and perhaps receive training on a remote instrument. Even if the researchers decide to visit the EMSL to conduct the actual experiment, they can meet people, understand procedures, and learn about the instrument before they arrive. Remote researchers must also find effective techniques for analyzing the data and consulting with co-researchers in writing up publications. Because scientific data are often complex and multidimensional, researchers will need to be able to confer with local researchers familiar with analysis of data from EMSL instruments."

The major difference between STIN models and HISTIN models is the extent to which the analyst assumes that the technological elements are socially constituted (and vice versa). In this Internet example, a STIN analysis might treat some elements, such as routers and servers as artifacts, but view the overall network of people, organizations, practices, and diverse devices as a STIN. In contrast, an HISTIN analysis would employ the heuristic that any artifact may be "opened up" to examine its social constitution; and any group may be examined to understand their co-constitution with various technologies. The HISTIN analyst doesn't open up all network nodes recursively; some are left unexamined. But an HISTIN analyst can ask how a router manages an activity like subnet addressing and how changes in the vendors' design teams' understanding and the vendors' marketing strategy will lead to different algorithms. Monteiro [24] carefully examined the debates for the new IP address standard and found that many of the Internet engineers were specially sensitive to the "organizationally structured" character of many components as they sought strategies for a smooth transition to a new standard. In short, HISTIN analysis are not just an esoteric social theory; they are part of the tacit understanding of many of the practicing computer scientists who have been developing Internet standards.

However, HISTIN analyses are more broadly applicable. Elsewhere, we have examined electronic journals and shown how their viability depends, not simply on the kinds of information processing features that the Layer Cake Model foregrounds, but also on their location in the STINs of their respective scientific communities [15,16]. Now we will apply an STIN analysis to a more complex set of scientific communication forums.

Our interests in framing an alternative to the Layer Cake Model of scientific communication forums are illustrated by the social interactions that energize collaboratory life that are briefly sketched in these accounts by Myers and others. These social and technical interactions seem to shape the work of collaboratories and their intellectual location in their own scientific fields. They are anomalies relative to the Layer Cake Model, but are central to the Sociotechnical Interaction Network models that we examine here.

**CASE STUDIES OF TWO COLLABORATORIES**
In this section we will discuss two different collaboratories to illustrate the usefulness of a STIN analysis.

**Case Study I: A Collaboratory in Materials Science**
The first case study is a brief account of a materials science collaboratory, MatterLab (pseudonymous). MatterLab is a small materials science collaboratory which is located within one of the U.S. national laboratories. It has a substantive mission to advance specific aspects of materials science. However, the initial goal of this project has been, according to their own materials, "to create a virtual space, accessible via the Internet," where materials scientists "and their colleagues, who are distributed across the nation or the world, can meet, talk, plan and also run their experiments." MatterLab is organized as a physical lab with several rooms, five electron microscopes, several computers, associated equipment and several desk-like work areas. It's on-line version is accessible via the WWW with a mixture of public spaces (general documents, Webcam shots of laboratory instrument consoles and their operators, Webcam shots of specific instruments and specimens) and password-controlled private spaces for discussion and documentation of specific experiments.

In our discussions, MatterLab's director stressed that the resident scientists who work "inside" a collaboratory are not just cognitive informants or passive technicians. Sometimes they act as evangelists, and attend conferences to recruit other scientists to use their facilities. When MatterLab was organized in 1996, some of its measuring equipment was very rare, and some scientists were drawn to the equipment. However, to advance materials science, MatterLab's scientists often have to reconfigure their equipment for a specific experiment. They believe that they must work as collaborators, and often expect to be co-authors of major papers that are based on their instruments and expertise.

During the late 1990s, the lab's cutting edge equipment has become much more widely available. The lab's director believes that his helpfulness and skill as a collaborator is the lure that draws and maintains collaborations today. MatterLab has to produce a set of interesting scientific results each year, and cannot expect to be funded for new cutting edge equipment every year. Thus MatterLab's staffs' abilities to maintain collaborations with outside scientists, and to tune their aging equipment for new experiments is a critical capability to maintain MatterLab as a going concern. Effectively tuning older equipment to produce new cutting edge science requires a deep understanding of both the materials science and the equipment. Thus, over time, the scientific imaginations and collaborative abilities of MatterLab's scientists becomes more important in sustaining their operation. Complex social interactions between the scientific "owners" of a collaboratory and outside scientists constitutes a collaboratory as much as its instrumentation.

In our terms, the collaboratory is constituted as a STIN that brings together people and equipment in ways that are not meaningfully separable for understanding "how collaboratories work." STINs highlight the importance of the character of the interactions between people, between people and equipment, and even between sets of equipment. Some of these interactions may involve direct participants, but the character of these interactions cannot be specified *a priori*. For example, the scientist who develops sustained trust is likely to work with a collaboratory very differently than one whom local experts are reluctant to work with. For example, "the collaboratory evangelist who turns out to be a zealot" can drive away other scientists, who could have been potential collaborators, away from a specific collaboratory. Further, the collaboratories are themselves nodes in a larger STIN of related (competing and cooperating) laboratories, funding relationships, publishing opportunities, and so on.

**Case Study II: Collaboratories in Experimental Particle Physics**
Experimental particle physics research is typically performed by large collaborations that include 50-1700 physicists. Before the inception of the Web, these physicists have developed Internet-based (and now, Web-based) communications forums in order to facilitate various aspects of collaboration, from communication of interim and final results to documentation of detector mechanics to public outreach to remote instrumentation. Most particle physics collaborations now have Web sites that are used by collaboration members for certain types of communication. Garrett and Ritchie [7] provide an overview of the evolution of particle physics Web-based collaboratories.

Although they do not conceptualize them as STINs, they implicitly use and exemplify STIN-based analysis.

The different collaboration Web sites are in various states of elaboration, and are used for different purposes by different groups. We will discuss one collaboratory – the Web site for the CONVEX collaboration at HEPLAB[6] as a final example. The use of the STIN approach in analyzing the Web site for the CONVEX collaboration in particle physics also highlights two other social relationships that have not yet been mentioned: the link between communication and work practice, and the articulation of communicative boundaries.

CONVEX is an experimental particle physics collaboration whose massive equipment is based at HEPLAB. It consists of physicists at over a dozen institutions studying certain charm (quark) phenomena. They gathered their data from their particle detector at HEPLAB. At the time of our field visit they were analyzing this data, but had not yet begun to publish their results.

*Link Between Communication and Work Practice*
Most physics collaboration Web sites are used extensively for documentary storage and retrieval. A few, including the CONVEX Web, are also used for remote instrumentation. Several physicists in the CONVEX collaboration developed a large suite of Web-based programs that allowed their collaborators to monitor remotely some of the instruments used in the data collection. During the six months of data taking, the CONVEX control room was staffed seven days a week, 24 hours a day by collaboration members. The work of the control room staff involves monitoring a number of activities, including the quality of several gasses in different parts of their detector, the quality of the positron beam entering the target area, and the number of particles that intercept the detector. Periodically, the gasses in a section of the detector degrade, and physicists have to replace gas bottles.

The remote instruments allowed some collaboration members who were offsite to participate more actively in data taking. They could observe a number of data displays which could be interpreted as indicating the quality of the positron beam, and could also observe indicators of the number of quark-related events that were being detected. The designers of the remote instrumentation wanted to enable the CONVEX collaborators who were not at HEPLAB at a given time to be able to observe the data taking and the experiment's progress.

However, on occasion remote collaborators seemed to know more about the beamline quality than the physicists who were actually working control rooms shifts at the time. Physicists on the CONVEX collaboration reported that remote collaborators would occasionally phone in to let them know that some parameter was out of specification on the beamline. The physicists in the control room may have been doing other work, such as changing gas bottles, when the beamline degraded. We sensed a significant ambivalence about these interventions by remote collaborators. On the one hand, they helped to keep the experiment on track and allowed the CONVEX collaboration to gather much more data than anticipated during their scheduled beamtime. On the other hand, the physicists in the control room seemed to feel 'caught short" and perhaps viewed as inattentive, when they were simply paying attention to another aspect of the experiments' complex operations.

*Sharing vs. Surveillance*
In the collaboratory literature, the primary mode of interpersonal interaction is sharing. The purpose of the collaboratories is to support the sharing of data, resources and instruments among collaborators. However, a simple example from the CONVEX Web site illustrates that other modes of interaction besides sharing can be seen: in this case, sharing the control room during data taking also slips into surveillance.

The remote instrumentation mentioned in the previous section eventually culminated in the placement of a digital video camera in the beam control room, so that remote users could even watch shift-workers during beam runs. However, after enough phone calls from other collaborators letting shift workers know that something was not right with the beam, many shift workers began to see the camera as a form of surveillance. Eventually, many collaborators in the control room turned the camera to face the ceiling, so remote users would simply see a blank screen!

*Sharing vs. Control: The Negotiated Boundaries Between Public and Private*
Even when sharing is the primary mode of interaction, authors and maintainers of communications forums may be concerned with what is public and what should not be made public. These boundaries between public and private are also structured and shaped by STINs. For example, many collaborations are in competition with one to four other collaborations. This competition leads some collaborations (such as KayBar at HEPLAB2[7] which is competing with a group at the Japanese lab, KEK) to take fairly elaborate measures to ensure that data is shared only within the collaboration (from segmenting of access to the use of Secure Sockets Layer (SSL) to ensure that the data channel itself is secure). Further, concern for professional reputation both individually and for the collaboration causes most, if not all, collaborations to have developed systems whereby only research results that have been officially "blessed" by the collaboration may be shared by the world on the collaboration Web site. Work in progress that has not yet been blessed by the collaboration is shared with other collaboration members via the Web site, but is password-protected against the outside world.

---

[6] CONVEX and HEPLAB are pseudonyms.

[7] KayBar and HEPLAB2 are also pseudonyms.

**Conclusions**

We have articulated a richer alternative, the STIN model, to the layer cake model of socio-technical systems as applied to collaboratories. We have examined how STIN models help understand important behavior in a materials science collaboratory and in a HEP collaboration. Like UARC/SPARC [28], the HEP collaboration existed prior to the development of online environments. In contrast, MatterLab was developed to help foster some new collaborations. Styles of scientific work differ across the sciences, and within them. For example, we expect different kinds of work practices and communications in small teams (MatterLab, UARC/SPARC) than in gigantic collaborations of 1700 physicists, such as ATLAS and CMS at CERN. However, we have found that STIN models help to highlight important behavior which is backgrounded or ignored with Layer Cake models.

One important consequence of adopting a STIN-based model is that it becomes clear that radical improvements in IT will not wash away the issues of sustainability and integration into a social world. For example, as the once-cutting-edge scientific instruments at MatterLab became more common elsewhere, the ability of MatterLab's scientists to be effective collaborators was more central to the collaboratories' scientific productivity. Social advances, such as developing workable co-authoring agreements are as important as having great technical environments.

Second, STIN-based analyses inject social analysis into all phases of planning, development, configuration, use, and evolution of a collaboratory, rather than merely at the beginning (in determining user "requirements"), and post-deployment (in determining the social "impacts") of the system. The examples of the MatterLab and HEP collaboratories help illustrate different types of social relationships foregrounded by an STIN-based analysis that are important to the use, sustainability, and evolution of collaboratories. The HEP collaboratories illustrate the extent to which working scientists are sensitive to selectively releasing information to others (and thus the importance of security protections as well as documentary and data sharing).

Third, the relevant STINs are not just constituted from CSCW tools and direct participants in a scientific teams. The weaker ties of competition with other teams that use better, lesser or just different instruments and research designs can influence the willingness of a given team to work with a specific collaboratory.

Fourth, the term "user" flattens the interactions of the scientists who wok in or with a specific collaboratory. STIN models portray them as social actors (or even interactors) whose work and communications are influenced by their locations in larger scale networks of scientists, funders, publishers, etc. The way that STIN models encourage CSCW researchers to move from relatively thinly depicted users to socially richer characterizations of people working and communicating in complex multivalent socio-technical networks that extend well beyond immediate workplaces and the most tightly coupled teams, may be most important.

All of these behaviors would be hard to anticipate from the Layer Cake Model of socio-technical systems. We suggest that future discussions of scientific communications forums, including collaboratories, should be informed by HISTIN models, or at least by STIN models. Their heuristic of seeking the social elements of technical formations and the technical supports for social life opens up important lines of inquiry to better understand these complex practices.

**ACKNOWLEDGEMENTS**

Funding was provided in part by NSF Grant #SBR-9872961 and with support from the School of Library and Information Science at Indiana University. This article benefited from helpful discussions with and comments from Dan Atkins, Blaise Cronin, Paul Dourish, Paul Edwards, Tom Finholt, Suzanne Iacono, Gary Olsen, Mark Napier, and Bill Turner. In addition, we are indebted to numerous scientists who have discussed their work practices with us.